\documentclass[runningheads]{llncs}
\usepackage[T1]{fontenc}
\usepackage{graphicx}
\usepackage{tabularx}
\usepackage{booktabs}
\usepackage{amsmath}
\begin{document}
\title{Combining Retrieval-Augmented Text Generation with LLMs for Reading Content Recommendations}
\titlerunning{RAG-LLM for Reading Recommendations}
%

\author{Sooyeon Kim \and Piotr S. Maciąg\orcidID{0000-0001-5486-7927}}
\authorrunning{Kim and Maciąg}

\institute{Institute of Computer Science, Warsaw University of Technology, Nowowiejska 15/19, 00-661 Warsaw, Poland\\
\email{piotr.maciag@pw.edu.pl}}
%
\maketitle              
\begin{abstract}
This work presents the design, implementation, and evaluation of a system for generating personalized reading content using Large Language Models (LLMs) combined with Retrieval-Augmented Generation (RAG). The proposed architecture consists of four modules—Input, RAG, Generation, and Judging—and enables users to specify both a question and a target reading content complexity. RAG is employed to retrieve relevant information from the Internet, enriching and grounding the content produced by three modern LLMs: Meta LLaMA~4~Scout, LLaMA~3.1--8B--Instant, and Google Gemma2--9B. Reading materials are generated using three prompting strategies (Chain-of-Thought, zero-shot, and few-shot), and the LLM-as-a-Judge module automatically evaluates answer quality and alignment with the desired readability level. Experimental results show that RAG consistently improves system performance across all models and prompting techniques, increasing relevance and particularly groundedness by up to 26--35 percentage points. Overall, the findings demonstrate that the RAG-augmented architecture effectively produces reading content tailored to user queries and desired textual complexity.
\keywords{RAG  \and LLMs \and Personalised Reading Content.}
\end{abstract}
\section{Introduction}
\label{sec:Intro}

Nowadays, Large Language Models (LLMs) are becoming increasingly popular for a wide range of tasks involving the generation of text or image content. In this article, we propose the use of LLMs as the backbone of a system for personalised reading content recommendation. According to a recent Eurostat survey~\cite{eurostat2025sustainable}, more than 50\% of students in Cyprus, Montenegro, North Macedonia, and Albania were classified as low achievers in at least one of the three assessed competency domains: reading, mathematics, or science. The same survey also reports that, although the proportion of low achievers in EU countries steadily decreased between 2006 and 2012, reaching its lowest values in 2015, it has since risen to record highs in 2022—26.2\%, 29.5\%, and 24.2\% for reading, mathematics, and science, respectively. One promising approach to counteracting this decline, particularly regarding reading skills, is the use of personalised reading materials, a strategy already widely adopted in domains such as marketing.

In this article, we present a personalised reading content generation system that combines LLMs with real-time web search through a Retrieval-Augmented Generation (RAG) component. To match the difficulty of the generated content to the user’s reading proficiency, we employ the Flesch--Kincaid Grade Level (FKGL) metric. FKGL assesses text complexity by combining average sentence length with word complexity based on syllable count, and maps the result onto U.S. school grade levels, enabling straightforward interpretation of text suitability for different reader groups (e.g., elementary, intermediate, or advanced). Unlike previous studies~\cite{attali2022interactive,drackert2025good}, which relied on human evaluation of generated content, our system employs an automated LLM-as-a-Judge framework, in which one LLM evaluates the outputs produced by another.

\textbf{Contributions.} We design a system for personalised reading recommendation that integrates LLMs with RAG module to produce text tailored to a user-defined target complexity level. This not only leverages LLMs for generating educational reading content but also incorporates real-time web search to retrieve relevant factual information and improve the quality of the generated texts.
Furthermore, we conduct a comprehensive set of experiments evaluating the effectiveness of different prompting strategies and LLMs in generating reading materials. The system is assessed using the LLM-as-a-Judge approach and evaluated with metrics such as relevance, groundedness, and a set of tolerance ranges measuring how closely the resulting FKGL complexity matches the target level.

\section{Methodology}
\label{sec:Methodology}

\subsection{The Designed Architecture}

\begin{figure}
    \centering
    \includegraphics[width=0.8\linewidth]{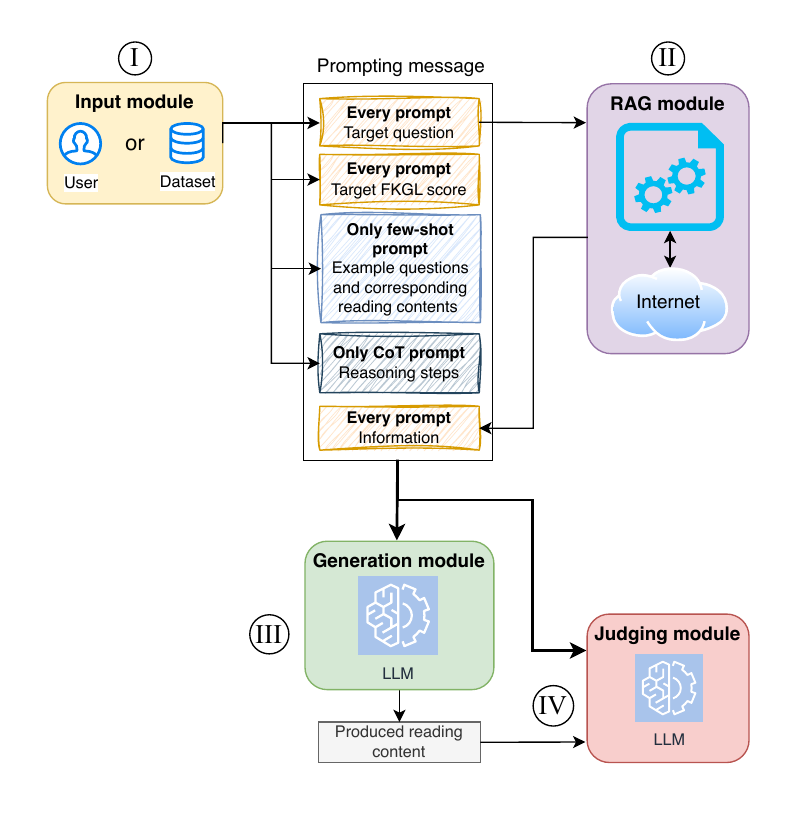}
    \caption{Designed architecture of reading content recommendation system. }
    \label{fig:Architecture}
\end{figure}

The designed system architecture is illustrated in Fig.~\ref{fig:Architecture}. It consists of four main modules: \textbf{Input}, \textbf{RAG}, \textbf{Generation}, and \textbf{Judging}. The \textbf{Input} module includes a user-provided target question (or one taken from the Natural Questions dataset described below), the target FKGL score of the resulting reading content, examples of questions and corresponding reading contents (used only in few-shot prompting), and a reasoning example (used only in CoT prompting). At the core of the system lies an LLM, which requires the selection of an appropriate prompting technique. In our experiments, we tested three distinct prompting methods:
\begin{enumerate}
    \item \textit{Zero-shot} — no examples of answers are provided to guide the LLM in generating the target response. The zero-shot technique was used in the initial experiments with LLMs as described, for example, by Radford et al. \cite{Radford2019-language-zeroshot}.
    
    \item \textit{Few-shot} — along with the target question, a few examples of questions and their corresponding answers are supplied. These examples help the LLM produce more accurate and contextually relevant answers \cite{Brown2020-Few-shot}. It is worth noting that such example demonstrations provided to an LLM do not update any model's weights. 
    
    \item \textit{Chain of Thought (CoT)} — this technique provides the target question together with explicit reasoning steps, encouraging the LLM to perform more structured and interpretable reasoning. As pointed out by Wei et al.~\cite{Wei2022-CoT}, CoT can significantly enhance an LLM’s ability to handle complex reasoning tasks, particularly in models with over 100 billion parameters. In our approach, we employed the \textit{CoT zero-shot prompting} version, in which the reasoning is not accompanied by additional examples but consists of specific steps that the LLM is encouraged to follow. Kojima et al.~\cite{Kojima2022-CoTZero} demonstrated this to be an effective prompting strategy.
\end{enumerate}

Subsequently, the \textbf{RAG} module retrieves relevant information from the web in response to the target question (e.g., a URL to a related Wikipedia page). 
Next, depending on the selected prompting technique, a \textit{prompting message} is constructed. The message is structured as follows:
\begin{itemize}
    \item for zero-shot prompting: \textit{(target question, target FKGL score, retrieved information)},
    \item for few-shot prompting: \textit{(target question, target FKGL score, retrieved information, example questions and answers)},
    \item for CoT prompting: \textit{(target question, target FKGL score, retrieved information, reasoning steps)}.
\end{itemize}   

The constructed prompting message is then passed to the \textbf{Generation} module. This module employs the LLM to \textit{generate the reading content}. In our experiments, we evaluated three different LLMs within this module:
\begin{enumerate}
    \item \textbf{Meta LLaMA 4 Scout}\footnote{\url{https://ai.meta.com/blog/llama-4-multimodal-intelligence/}} — a model consisting of 109 billion parameters, of which 17 billion are so-called active parameters, leveraging a mixture-of-experts architecture composed of 16 experts. Each expert represents a specialized sub-network within the model. Depending on the input, only a subset of experts is activated, significantly improving computational efficiency. Furthermore, the model features an exceptionally large context window, theoretically capable of processing up to 10 million input tokens. LLaMA 4 Scout was released on April 5, 2025, with a knowledge cut-off date of April 2024. It is a multimodal and multilingual model, accepting both text and image inputs.

    \item \textbf{LLaMA 3.1-8B-Instant}\footnote{\url{https://ai.meta.com/blog/meta-llama-3-1/}} — a smaller model compared to LLaMA 4 Scout, containing 8 billion parameters. Unlike Scout, it does not employ expert sub-networks, meaning that all parameters are active for any given input. The model officially supports eight natural languages, accepts only text input, and has a knowledge cut-off date of December 2023.  

    \item \textbf{Google Gemma2-9B} \cite{Team2024-gemma} — a lightweight model consisting of 9 billion active parameters (similar to LLaMA 3.1-8B-Instant). It was trained on a mixture of web documents, programming code, and mathematical formulae. The exact knowledge cut-off date has not been publicly disclosed.
\end{enumerate}

Finally, the \textbf{Judging} module evaluates the generated output based on both the prompting message and the produced text. In our approach, this module also employs the LLM component. Specifically, OpenAI’s GPT-4.1 model is used as the evaluator. The model was released on April 14, 2025, with a knowledge cut-off date of June 1, 2024. The use of an \textit{LLM-as-a-Judge} has become increasingly popular in various tasks, such as data annotation~\cite{Gu2024-llmjudge-survey}. Using the MT-Bench question benchmark, and generating corresponding answers with various LLMs, Zheng et al.~\cite{zheng2023judging} demonstrated that GPT-4, when used as a judge, achieves up to 80\% agreement with human experts and volunteers in answer evaluation.

The evaluation process employed by us considers not only the difference between the actual FKGL score of the generated content and the target score, but also three additional metrics: \textit{Relevance} and \textit{Groundedness}. These metrics are discussed in detail in subsection~\ref{subsec:EvaluationProcedure}. 

\subsection{Components of the Software Implementation}

Several software tools were used to implement the system. Specifically:
\begin{itemize}
    \item \textbf{LangChain}\footnote{\url{https://www.langchain.com}} is a platform based on Python that allows to build applications powered by LLMs. In our project, LangChain is used as a backbone of the system, orchestrating control flow between different modules and to compose prompting messages. 
    
    
    \item \textbf{LangSmith}\footnote{\url{https://www.langchain.com/langsmith/observability}} is a monitoring tool that enables real-time tracing of the execution flow in LLM-based systems through a graphical user interface (GUI). LangSmith allows users to track various metrics, alerts, and error rates of the designed system.
    
    
    \item \textbf{Tavily}\footnote{\url{https://www.tavily.com}} - it is an API that allows to perform RAG operation over the Internet. In our system, the RAG modules uses Tavily to search for the relevant information (such as link to websites) given the target user/dataset question.
    
    \item \textbf{GroqCloud}\footnote{\url{https://console.groq.com/home}} in our system, the three tested LLMs (Meta AI’s LLaMA 4 Scout, LLaMA 3.1-8B-Instant, and Gemma2-9B) used by the Generation module are hosted in this platform.
    
    \item \textbf{Microsoft Azure Platform}\footnote{\url{https://azure.microsoft.com/}} - the Judging module uses the OpenAI’s GPT-4.1 hosted in this platform. 
\end{itemize}

\subsection{The Used Dataset}
\label{subsec:Dataset}

To systematically evaluate the system, we used the \textit{Natural Questions}\footnote{\url{https://ai.google.com/research/NaturalQuestions/download}} dataset of open-domain questions collected from anonymized real queries to Google Web Search. The dataset covers a wide range of topics, from everyday questions to academic subjects, making it highly suitable for evaluating RAG systems. The full dataset consists of 7,830 questions accompanied by answers, which are typically links to relevant Wikipedia pages. 

For our experiments, we extracted a subset of 30 questions from the full dataset. While this number may appear small, it is consistent with other studies evaluating various aspects of text generation using LLMs (for example, Zheng et al.~\cite{zheng2023judging} tested LLM accuracy using only 80 questions). 

Each selected question was assigned a target FKGL score representing the desired reading complexity of the generated content. The assigned scores ranged from 2 (least complex) to 12 (most complex). Examples of selected questions and their corresponding target FKGL scores include:
\begin{itemize}
    \item \textit{What is the summary of "Magnus Chase and the Gods of Asgard"?} — Target FKGL score: 12.
    \item \textit{What is the plot of "Bendy and the Ink Machine"?} — Target FKGL score: 8.
    \item \textit{How many breeds of pigs are there in the UK?} — Target FKGL score: 7.
\end{itemize}

\subsection{The Evaluation Procedure}
\label{subsec:EvaluationProcedure}

In our approach, we evaluated the designed system using three measures. First, we aimed to verify how well a given prompt (and the corresponding target question) aligns with the resulting output. To achieve this, we selected two metrics that are widely used when assessing LLM-based systems, particularly those that employ the Retrieval-Augmented Generation (RAG) technique.  
A short description of each metric is provided below:

\begin{enumerate}
    \item \textbf{Relevance} – measures the degree to which the generated answer aligns with the user’s original question.
    \item \textbf{Groundedness} – assesses whether the produced answer is strictly based on the information supplied to the model.
\end{enumerate}

These metrics, together with the judgement criteria provided to the \textbf{LLM-as-a-Judge} module, are summarised in Table~\ref{Table:metrics}. It is possible that, for a given question and the generated answer, the evaluation may indicate that the output is relevant, yet not grounded.

\begin{table}[h!]
\centering
\caption{Evaluation metrics and criteria used in the system.}
{
\begin{tabularx}{\textwidth}{cX}
\toprule
\textbf{Metric} & \textbf{Criteria provided to the \textit{LLM-as-a-Judge} module} \\ 
\midrule

Relevance & 
Ensure the answer covers all essential requirements stated in the question (facts, figures, concepts, etc.).\par
Ensure the answer helps the user solve the problem or improves their understanding.\par
Ensure the answer directly correlates with the question. \\ 
\midrule


Groundedness &
Ensure that at least one retrieved fact is linked to the answer.\par
Ensure that the answer is logically supported by the provided facts.\par
Ensure that the answer does not contradict any of the retrieved facts. \\ 
\bottomrule
\end{tabularx}
}
\label{Table:metrics}
\end{table}

Furthermore, to evaluate the readability of the generated output, we employed the FKGL score \cite{kincaid1975readability}, calculated according to Eq.~\ref{Eq:FKGL}. The FKGL score corresponds to a U.S.\ grade level and reflects the number of years of education typically required to understand a given text.

{
\begin{equation}
\text{FKGL}
= 0.39 \times \frac{\text{total\_words}}{\text{total\_sentences}}
+ 11.8 \times \frac{\text{total\_syllables}}{\text{total\_words}}
- 15.59
\label{Eq:FKGL}
\end{equation}
}

Instead of only checking whether the calculated FKGL score of the generated answer exactly matches the target FKGL score provided by the user, we also verify whether the resulting score falls within a reasonable tolerance range relative to the target value. The tolerance ranges E1--E6 defined for this purpose are presented in Table~\ref{Table:fkglCalc}. For a given range, if the resultant FKGL and the target FKGL satisfy the corresponding calculation criteria, the generated answer is considered correct (True); otherwise, it is marked as incorrect (False).

\begin{table}[h!]
\centering
{
\caption{Tolerance ranges for the FKGL score ($\lfloor x \rfloor$ denote the mathematical floor function of real number $x$).}
\begin{tabularx}{\textwidth}{cX}
\toprule
\textbf{Range} & \textbf{Calculation criteria} \\ 
\midrule

E1 & If $(\lfloor \text{resultant FKGL} \rfloor = \text{target FKGL})$ return $True$; else $False$.\\ 
\midrule

E2 & If $(\text{target FKGL} - 1 \leq \lfloor \text{resultant FKGL} \rfloor \leq \text{target FKGL} + 1)$ return $True$; else $False$. \\ 
\midrule

E3 & If $(\text{target FKGL} - 2 \leq \lfloor \text{resultant FKGL} \rfloor \leq \text{target FKGL} + 2)$ return $True$; else $False$. \\ 
\midrule

E4 & 
If $(\text{target FKGL} \leq 6 \ \text{AND}\ \lfloor \text{resultant FKGL} \rfloor \leq 6)$ return $True$; \par
else if $(7 \leq \text{target FKGL} \leq 11 \ \text{AND}\ 7 \leq \lfloor \text{resultant FKGL} \rfloor \leq 11)$ return $True$; \par  
else if $(\text{target FKGL} \geq 12 \ \text{AND}\ \lfloor \text{resultant FKGL} \rfloor \geq 12)$ return $True$; \par
else return $False$. \\ 
\midrule

E5 & 
If $(\text{target FKGL} \leq 6 \ \text{AND}\ \lfloor \text{resultant FKGL} \rfloor \leq 7)$ return $True$; \par
else if $(7 \leq \text{target FKGL} \leq 11 \ \text{AND}\ 6 \leq \lfloor \text{resultant FKGL} \rfloor \leq 12)$ return $True$; \par  
else if $(\text{target FKGL} \geq 12 \ \text{AND}\ \lfloor \text{resultant FKGL} \rfloor \geq 11)$ return $True$; \par
else return $False$. \\ 
\midrule 

E6 & 
If $(\text{target FKGL} \leq 6 \ \text{AND}\ \lfloor \text{resultant FKGL} \rfloor \leq 8)$ return $True$; \par
else if $(7 \leq \text{target FKGL} \leq 11 \ \text{AND}\ 5 \leq \lfloor \text{resultant FKGL} \rfloor \leq 13)$ return $True$; \par  
else if $(\text{target FKGL} \geq 12 \ \text{AND}\ \lfloor \text{resultant FKGL} \rfloor \geq 10)$ return $True$; \par
else return $False$. \\ 

\bottomrule
\end{tabularx}
}
\label{Table:fkglCalc}
\end{table}


\section{Results}
\label{sec:Results}

\begin{figure}
    \centering
    \includegraphics[width=0.85\linewidth]{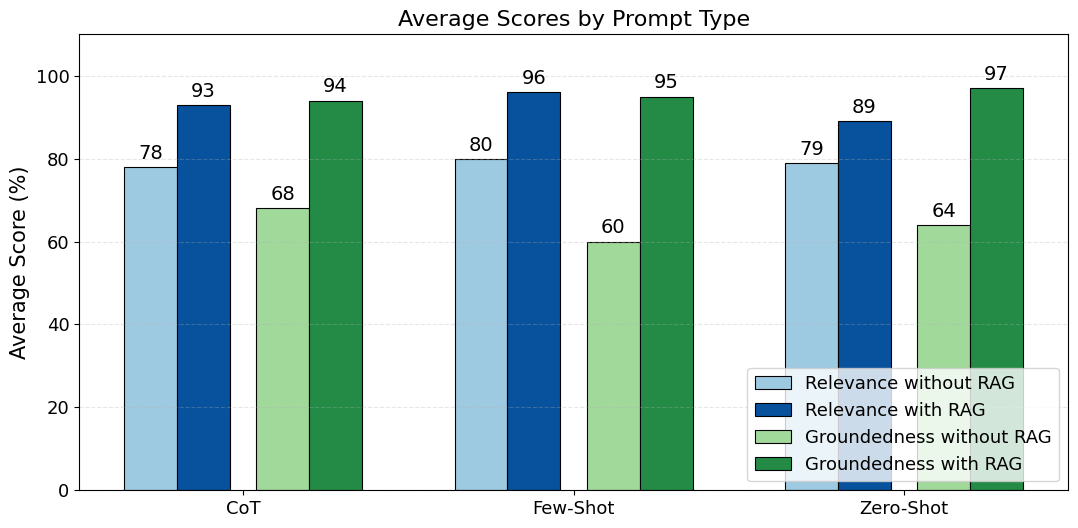}
    \caption{Obtained average relevance and groundedness for different prompt techniques with and without using the RAG module of the system.}
    \label{fig:results-prompt}
\end{figure}

We first present results for relevance and groundedness for the different prompting techniques applied in the system (zero-shot, few-shot, and CoT), as well as for the selected LLMs. Subsequently, we report the results of the FKGL readability evaluation. Fig.~\ref{fig:results-prompt} shows the average relevance and groundedness scores for the three prompting techniques, both with and without the RAG module. The averages were computed across the answers to all 30 questions selected from the dataset.

For each prompting technique, the use of the RAG module substantially improves the average values of both metrics. Overall, the largest gains in both relevance and groundedness are observed for the few-shot. The particularly strong improvements in the groundedness metric indicate that the RAG module provides much better factual support, effectively reducing model hallucinations.

\begin{figure}
    \centering
    \includegraphics[width=0.85\linewidth]{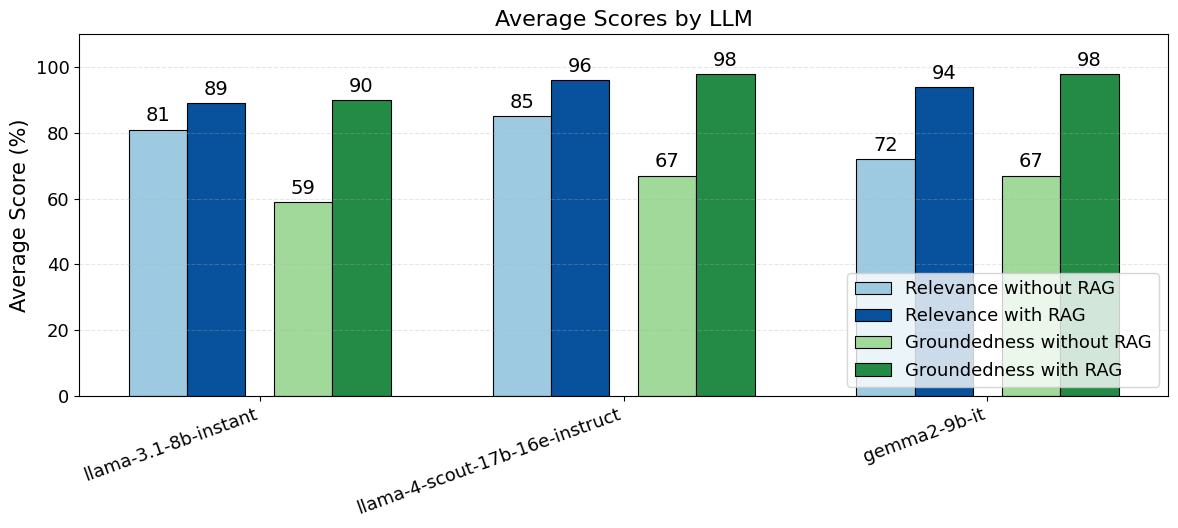}
    \caption{Obtained average relevance and groundedness for different selected LLMs with and without using the RAG module of the system.}
    \label{fig:results-model}
\end{figure}

Fig.~\ref{fig:results-model} presents a similar comparison of average relevance and groundedness, but across different LLMs producing the reading content. All models exhibit a 31-point improvement in groundedness when RAG is enabled, which again provides strong evidence that hallucinations primarily arise from missing contextual information. Consequently, regardless of the chosen model, the use of RAG consistently improves the obtained results. The impact of RAG is systematic and consistent across all prompting techniques and all evaluated models.


\begin{figure}
    \centering
    \includegraphics[width=0.85\linewidth]{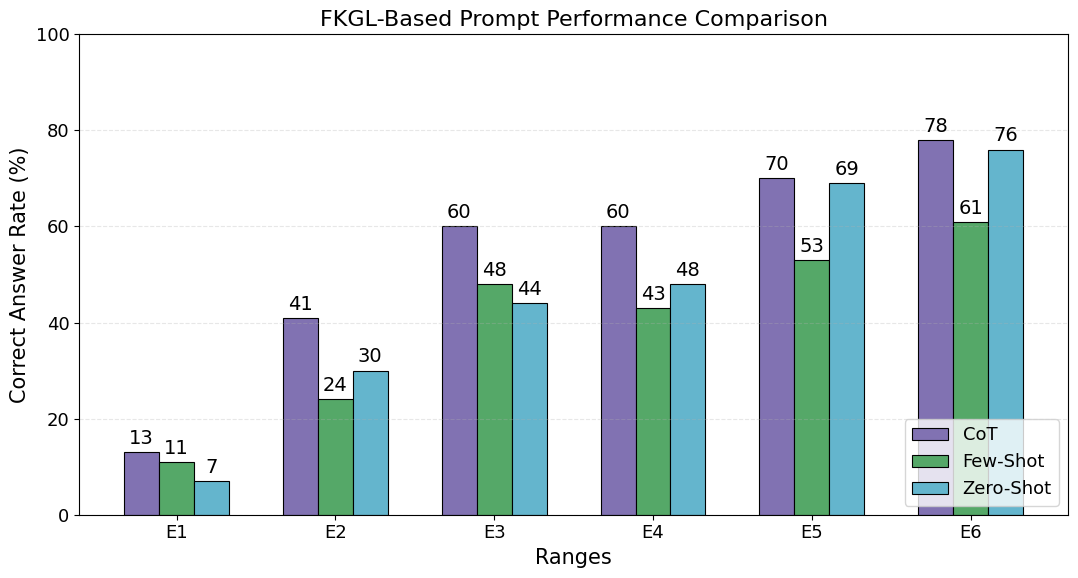}
    \caption{Evaluation of the FKGL metric for different prompting techniques and tolerance ranges.}
    \label{fig:fkgl-prompts}
\end{figure}

\begin{figure}
    \centering
    \includegraphics[width=0.85\linewidth]{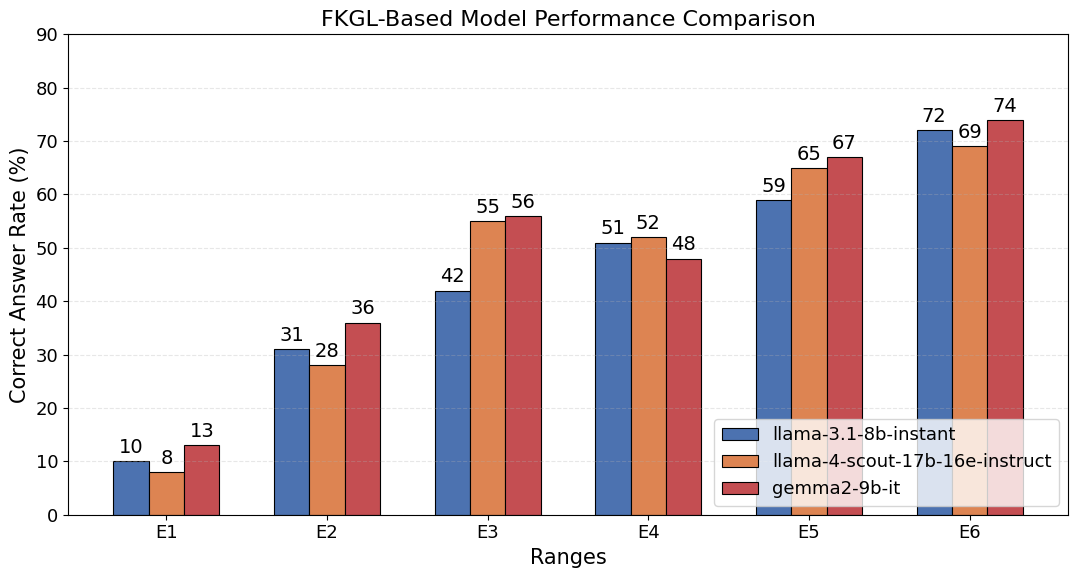}
    \caption{Evaluation of the FKGL metric for different models and tolerance ranges.}
    \label{fig:fkgl-models}
\end{figure}

In Figs.~\ref{fig:fkgl-prompts} and~\ref{fig:fkgl-models}, we present the evaluation of the generated answers using the proposed FKGL tolerance ranges, applied across different prompting techniques and LLMs, respectively. In these comparisons, the RAG augmentation is always enabled. Across all tolerance ranges, the CoT prompting technique yields the highest scores, generally followed by the zero-shot prompting. With regard to model selection, the \textit{gemma2-9b-it} model typically achieves the best results. As indicated by the values obtained for the strictest tolerance range~(E1), regardless of the chosen prompting technique or model, the system is able to generate an answer whose FKGL score exactly matches the target score for only about 10\% of the questions. However, when the tolerance range is relaxed to E2 or E3, the performance improves considerably. In particular, for nearly half of all generated answers, the resultant FKGL score matches the target FKGL score within the E3 tolerance range, irrespective of the prompting technique or selected model.


To summarize the achieved results we conclude that across all experiments, the use of the RAG module substantially improved the quality of the generated answers. For all prompting techniques and LLMs, RAG consistently increased both relevance and groundedness, with gains in groundedness reaching up to 26--35 percentage points. Differences between prompting strategies were more visible without the use of the RAG module; however, once RAG was applied, their performance converged, with CoT achieving the highest overall scores. Similarly, larger LLMs outperformed the smaller model in the non-RAG setting, but these differences largely disappeared when RAG was enabled. FKGL evaluation further showed that exact matching of the target readability level E1 occurred in only about 10\% of the cases, while nearly half of all generated answers matched the target within the more permissive E3 range.

\section{Related Work}
\label{sec:RelatedWork}

LLMs have recently established themselves as powerful and versatile tools for a broad range of natural language and image processing tasks. A comprehensive review of advancements in LLM design, including architectural developments, training methods, word-embedding techniques, and historical evolution, is provided in~\cite{kumar2024large}. The design of effective prompting techniques has attracted substantial research attention, as prompting is often a more efficient alternative to model fine-tuning, which may require significant computational resources~\cite{zhao2023survey}. Zhao et al.~\cite{zhao2023survey} classify prompting techniques into two main categories: Chain-of-Thought (CoT) prompting and In-Context Learning (ICL). The latter includes approaches such as few-shot prompting, also used in our work. As noted in several studies (e.g.,~\cite{han2023understanding}), the effectiveness of ICL can vary substantially depending on the quality of the provided examples and the templates used to present them. CoT-based prompting aims to enhance the reasoning capabilities of LLMs by supplying explicit step-by-step examples, which also offer greater interpretability of the model’s internal reasoning. 

Other researchers have also explored the use of LLMs for generating educational reading content. Drackert et al.~\cite{drackert2025good} focused specifically on evaluating reading texts in German. The authors compared high-stakes assessment texts with texts generated by ChatGPT~3.5 and ChatGPT~4, using a variety of linguistic features as well as evaluations provided by human experts. Attali et al.~\cite{attali2022interactive} employed the ChatGPT-3 model to generate short reading passages using a few-shot prompting technique. In contrast to our work, the quality of the generated content in their study was assessed exclusively by human experts.

\section{Summary}
\label{sec:Summary}

In this work, we designed, implemented, and evaluated a system for generating personalized reading content using LLMs combined with Retrieval-Augmented Generation (RAG). The system consists of four modules—Input, RAG, Generation, and Judging—and allows users to submit a question together with a target FKGL readability level. Across three LLMs (Meta LLaMA~4~Scout, LLaMA~3.1--8B--Instant, and Google Gemma2--9B) and three prompting strategies (CoT, zero-shot, and few-shot), RAG consistently improved output quality, increasing relevance and especially groundedness by up to 26--35 percentage points, indicating reduced hallucinations and stronger factual support. Although experiments used 30 Natural Questions items, the improvement is expected to transfer to larger sets because it primarily results from the same retrieval mechanism—supplying question-relevant evidence—rather than from question-specific memorization; however, larger-scale evaluation may refine the estimated gain under noisier retrieval and broader topic diversity.

FKGL evaluation showed that exact matching of the target level (E1) is difficult (about 10\% accuracy), largely because FKGL depends on surface features (sentence length and syllable-based word complexity) that are only indirectly controlled during generation and can shift with small paraphrases. Moreover, retrieval may introduce necessary multi-syllabic terminology, which improves groundedness but can increase FKGL, creating a trade-off under strict tolerance. Regarding deployment, the main cost drivers are web retrieval calls and multiple LLM invocations (generation and judging); practical systems can mitigate these via caching retrieval results, limiting retrieved documents, and running the judge selectively (e.g., only when FKGL deviation is high or for higher-stakes use cases).

\begin{credits}
\subsubsection{\ackname} The second author was supported by the Warsaw University of Technology Research University - Excellence Initiative program [Grant number CPR-IDUB/288/Z01/POB3/2024].

\subsubsection{\discintname}
The authors have no competing interests to declare that are
relevant to the content of this article.

\subsubsection{The use of AI tools}
The LLM model (ChatGPT-5) was used to polish writing of the manuscript, mainly including correction of grammar mistakes. 

\end{credits}



\bibliographystyle{splncs04}
\bibliography{my-references}

\end{document}